\documentstyle[preprint,prl,aps]{revtex}
\begin{document}
\def\sn2{$\sin^22\theta$}
\def\dm2{$\Delta m^2$}
\def\ch2{$\chi^2$}
\def\ltap{\ \raisebox{-.4ex}{\rlap{$\sim$}} \raisebox{.4ex}{$<$}\ }
\def\gtap{\ \raisebox{-.4ex}{\rlap{$\sim$}} \raisebox{.4ex}{$>$}\ }
\draft
\begin{titlepage}
\preprint{\vbox{\baselineskip 10pt{
\hbox{IASSNS -- AST 96/11}
\hbox{Ref. SISSA 89/95/EP}
%\hbox{hep -- ph/95?}
\hbox{February 1996}}}}
\vskip -0.4cm
\title{ \bf ON THE MSW $\nu_e \rightarrow \nu_s$ TRANSITION 
          SOLUTION OF THE SOLAR NEUTRINO PROBLEM}
\author{P. I. Krastev\footnote{Also at:
Institute of Nuclear Research and Nuclear Energy, Bulgarian Academy
of Sciences, BG--1784 Sofia, Bulgaria.}}
\address{School of Natural Sciences,
Institute for Advanced Study, \\
Princeton, New Jersey 08540,}
\vglue -0.4cm
%%\author{and}
%%\vglue 0.3cm
\author{S.T. Petcov$^{*~a,b)}$ and L. Qiuyu$^{a)}$}
\address{$^{a)}$Scuola Internazionale Superiore di Studi Avanzati, and\\ 
$^{b)}$Istituto Nazionale di Fizica Nucleare, Sezione di Trieste, I-34013 
Trieste, Italy}
\maketitle
\begin{abstract}
\begin{minipage}{5in}
\baselineskip 16pt
We study the stability of the two--neutrino 
MSW solution of the solar neutrino problem,
corresponding to solar $\nu_e$ transitions  
into sterile neutrino, $\nu_e \rightarrow \nu_s$, 
with respect to changes of the
total fluxes of $^{8}$B and $^{7}$Be neutrinos, $\Phi_{{\rm B}}$ and
$\Phi_{{\rm Be}}$. For any value of $\Phi_{{\rm Be}}$ from the interval
$0.7\Phi^{{\rm BP}}_{{\rm Be}}\leq \Phi_{{\rm Be}}
\leq 1.3\Phi^{{\rm BP}}_{{\rm Be}}$
(for $\Phi_{{\rm Be}} = 0.7\Phi^{{\rm BP}}_{{\rm Be}}$) the 
$\nu_e \rightarrow \nu_{s}$ MSW transitions 
provide at 95\% C.L. a description of the existing solar neutrino data
for $0.40~(0.39)~\Phi^{{\rm BP}}_{{\rm B}} \ltap \Phi_{{\rm B}} \ltap
36~(40)\Phi^{{\rm BP}}_{{\rm B}}$, 
$\Phi^{{\rm BP}}_{{\rm B}}$ and $\Phi^{{\rm BP}}_{{\rm Be}}$ being the fluxes
in the solar model of Bahcall--Pinsonneault from 1992. The corresponding 
allowed regions of values of the parameters \dm2 and \sn2, characterizing the
solar neutrino transitions, are derived. The physical implications of the found
MSW $\nu_e \rightarrow \nu_s$ solutions for the future solar neutrino
experiments are considered as well.
\end{minipage}
\end{abstract}
\end{titlepage}
\newpage

\hsize 16.5truecm
\vsize 24.0truecm
\def\dm{$\Delta m^2$\hskip 0.1cm }
\def\dmf{$\Delta m^2$}
\def\sn{$\sin^2 2\theta$\hskip 0.1cm }
\def\snf{$\sin^2 2\theta$}
\def\trna{$\nu_e \rightarrow \nu_a$}
\def\trnm{$\nu_e \rightarrow \nu_{\mu}$}
\def\trns{$\nu_e \leftrightarrow \nu_s$}
\def\trnat{$\nu_e \leftrightarrow \nu_a$}
\def\trnmt{$\nu_e \leftrightarrow \nu_{\mu}$}
\def\trne{$\nu_e \rightarrow \nu_e$}
\def\trnst{$\nu_e \leftrightarrow \nu_s$}
\font\eightrm=cmr8
\def\aprle{\buildrel < \over {_{\sim}}}
\def\aprge{\buildrel > \over {_{\sim}}}
\renewcommand{\thefootnote}{\arabic{footnote}}
\setcounter{footnote}{0}
%%\line{ }
\vglue 1.5cm
\leftline{\bf 1. INTRODUCTION}
\vskip 0.3cm
\indent Recent developments in the domain of the solar neutrino problem [1,2]
(see also the review article [3]) brought two new important
elements. First, the spectrum of solar model predictions [4--9] for
the flux of $^8{\rm B}$ neutrinos, $\Phi_{{\rm B}}$, now spans the
entire range of estimated 99.9\% C.L. uncertainties [2]. The central value of
$\Phi_{{\rm B}}$ derived in the ``low'' flux model [9] is smaller
approximately by the factors 2.0 and 2.3 than the central values in the
``high'' flux models of refs. [4] and [8], respectively
\footnote{Extensive discussions of the possible sources of
uncertainties in the solar model predictions for $\Phi_{{\rm B}}$ are
given in refs. [2--5,8--11].}. Second, it was found that the upper
limits on the $^{7}$Be neutrino flux, $\Phi_{{\rm Be}}$, which can be
inferred from the existing solar neutrino data [1,12--14] (see also
ref. [15]), are significantly lower [16] than the values predicted by
the solar models.  The predictions [4--9] for $\Phi_{{\rm Be}}$ vary
only by $\sim$25\% (from $\Phi_{{\rm Be}} = 4.20 \times
10^{9}~\nu_e$/cm$^{2}$/sec in ref. [9] to $\Phi_{{\rm Be}} = 5.18
\times 10^{9}~\nu_e$/cm$^{2}$/sec in ref. [8]).  No plausible
astrophysical and/or nuclear physics explanation of the indicated
beryllium neutrino deficit has been proposed so far, which represents
a major new aspect of the solar neutrino problem.

   Having in mind the spread in the current solar model predictions for 
$\Phi_{{\rm B}}$ and $\Phi_{{\rm Be}}$, we investigate in the present article 
the stability of the MSW solution of the solar neutrino problem [17--19] 
with solar $\nu_e$ transitions into sterile neutrino $\nu_s$ (see,
e.g., ref. [19]), with respect to variations of the values of total fluxes of 
$^{8}$B and $^{7}$Be neutrinos. Assuming that the $^{7}$Be neutrino flux has a 
value in the interval $0.7\Phi^{{\rm BP}}_{{\rm Be}} \leq \Phi_{{\rm Be}}
\leq 1.3\Phi^{{\rm BP}}_{{\rm Be}}$, where $\Phi^{{\rm BP}}_{{\rm Be}}$ is the
flux predicted in the reference solar model of Bahcall -- Pinsonneault [4],
we determine the range of values of the $^{8}$B neutrino
flux, for which the results of the solar neutrino experiments can be described
in terms of two--neutrino MSW transitions of the solar neutrinos into a
sterile neutrino, $\nu_e \rightarrow \nu_{s}$.
Similar analyses for the MSW solution with solar $\nu_e$ transitions into an
active neutrino [18,19], $\nu_e \rightarrow \nu_{\mu (\tau)}$, and for the 
different possible vacuum oscillation solutions [20,21,22] 
\footnote{More complete lists of 
references on the vacuum oscillation solution of the solar neutrino problem
can be found in refs. [21,22].}, were performed respectively in refs.
[23,24] and [22] (see also [11,25]). We find that the MSW
$\nu_e \rightarrow \nu_{s}$ transitions provide at 95\% C.L. a solution
of the solar neutrino problem for any value of $\Phi^{{\rm BP}}_{{\rm Be}}$
from the interval
$0.7\Phi^{{\rm BP}}_{{\rm Be}} \leq
\Phi_{{\rm Be}} \leq 1.3\Phi^{{\rm BP}}_{{\rm Be}}$
(for $\Phi_{{\rm Be}} = 0.7\Phi^{{\rm BP}}_{{\rm Be}}$) if
$\Phi_{{\rm B}}$ lies within the remarkably wide interval
$0.40~(0.39)~\Phi^{{\rm BP}}_{{\rm B}} \ltap \Phi_{{\rm B}} \ltap
36~(40)\Phi^{{\rm BP}}_{{\rm B}}$, $\Phi^{{\rm BP}}_{{\rm B}}$ being the 
$^{8}$B neutrino flux in the reference model [4].
The corresponding allowed regions of values of the parameters \dm and \sn,
characterizing the transitions, are derived. The physical implications of the
found ``low'' and ``high'' $\Phi_{{\rm B}}$ 
MSW $\nu_e \rightarrow \nu_{s}$ transition solutions for the future solar 
neutrino experiments are considered as well.

   We shall use the latest published data from the solar neutrino
experiments [1,12--14] in the present analysis:
$$\bar{\rm R}({\rm Ar}) = (2.55~\pm~0.25)\hskip 0.2cm {\rm SNU},~~~\eqno(1)$$
%%$(2.55 \pm 0.17 \pm 0.18)\hskip 0.2cm {\rm SNU}$
%%(first error is statistical, the second is systematic, and the error in the
%%second equation is obtained by adding in quadratures the two errors.)
$$\bar\Phi^{{\rm exp}}_{{\rm B}} = (2.89~\pm~0.42) \times 10^{6}~{\rm
cm^{-2}sec^{-1}},~~\eqno(2)$$ $$\bar {\rm R}_{\rm GALLEX}({\rm Ge}) =
(77.1~^{+9.60}_{-10.1})
\hskip 0.2cm {\rm SNU},~~\eqno(3)$$
$$\bar {\rm R}_{{\rm SAGE}}({\rm Ge}) = (69~\pm~13)\hskip 0.2cm {\rm
SNU},~~\eqno (4)$$
\noindent where $\bar{\rm R}$(Ar), and $\bar{\rm R}_{\rm GALLEX}({\rm Ge})$ and
$\bar{\rm R}_{\rm SAGE}({\rm Ge})$, are respectively the average rates of
$^{37}$Ar and $^{71}$Ge production by solar neutrinos observed in the
experiments of Davis et al. [1], and GALLEX [13] and SAGE [14], and
$\bar{\Phi}^{{\rm exp}}_{{\rm B}}$ is the flux of $^{8}$B neutrinos measured
by the Kamiokande collaborations [12]. In eqs. (1) -- (4)
the quoted errors represent the added in quadratures statistical (1 s.d.) and
systematic errors. 
\vskip 0.3cm
\leftline{\bf 2. MSW $\nu_e \rightarrow \nu_s$ TRANSITION SOLUTIONS: GENERAL
PROPERTIES}
\vskip 0.2cm
  It is convenient to introduce the parameters
$${\rm f_{B}} \equiv {{\rm \Phi_{B}}\over {\rm \Phi_{B}^{BP}}} \geq 0,~~~~~
  {\rm f_{Be}} \equiv {{\rm \Phi_{Be}}\over {\rm \Phi_{Be}^{BP}}} \geq 0,
             ~~~\eqno(5)$$
\noindent in terms of which we shall describe the possible
deviations of the $^{8}$B and $^{7}$Be neutrino fluxes, ${\rm
\Phi_{B}}$ and ${\rm \Phi_{Be}}$, from the values they have in the
reference model [4]. The fluxes ${\rm \Phi_{B}}$ and ${\rm \Phi_{Be}}$
predicted in the models [4,5,8,9] correspond, respectively, to $~{\rm
f_{B}}=~$ 1.0; 0.78; 1.14; 0.50, and $~{\rm f_{Be}}=~$1.0; 0.89; 1.06;
0.86.

   The Kamiokande data imposes at 99.73\% (95\%) C.L. the following 
limits on the values ${\rm f}_{{\rm B}}$ can possibly have [22]:
$${\rm f_{B}} \gtap 0.30~(0.37),~~~~~\eqno (6)$$
$${\rm f_{B}} \ltap 4.8~(4.3).~~~~~\eqno (7)$$
\noindent The lower limit (6) holds [22,11] in the case of solar two--neutrino
transitions or oscillations into active ($\nu_{\mu (\tau)}$ or
$\bar {\nu}_{\mu (\tau)}$) or sterile ($\nu_{s}$) neutrino, as well as for
transitions or oscillations of $\nu_e$ involving more than two neutrinos 
(active and/or sterile). The limit (6) is universal: it does not depend on the 
type of possible transitions, and on the specific mechanism responsible for 
them. 
 
  Contrary to the lower limit (6), the upper limit (7) is valid only  
for two--neutrino solar $\nu_e$ transitions or oscillations into an active
neutrino $\nu_{\mu (\tau)}$ or $\bar {\nu}_{\mu (\tau)}$ [22,11]. It does not 
apply, in particular, to transitions of the solar $\nu_e$ into sterile 
neutrino(s). 
%The limit (7) is also universal: it does not depend on the type of the 
%mechanism which causes the oscillations or the transitions.

   Owing to the specific dependence of the MSW two--neutrino transition
probability on the neutrino energy E it is possible to derive, using the 
Kamiokande or the Cl--Ar data, a somewhat constraining upper bound on 
$~{\rm f_{B}}$ in the case of $\nu_{e} \rightarrow \nu_{s}$ MSW transitions.
For $\sin^22\theta \gtap 4\times 10^{-3}$ [26] the relevant (averaged over the
corresponding region of $\nu_e$ production in the Sun, etc.) 
MSW $\nu_e$ survival probability, ${\rm \bar{P}_{MSW}(E)}$, satisfies [17] 
min ${\rm \bar{P}_{MSW}(E)} \cong \sin^2\theta~$
for certain interval of values of E when the transitions are adiabatic.
Utilizing this property of ${\rm \bar{P}_{MSW}(E)}$ and taking into account 
that the contribution due to the $^{8}$B neutrinos in the argon production rate 
${\rm \bar{R}(Ar)}$ in the Cl--Ar experiment cannot exceed the measured value 
of ${\rm \bar{R}(Ar)}$, eq. (1), one finds:
$${\rm MSW},~\nu_{e} \rightarrow \nu_{s}~:~
~~~~~~~~~~~~~~~{\rm f_{B}} < 67~(61).
~~~~~~~~~~~~~~~~~~~~~~~~~~~~~~\eqno (8)$$
\noindent This bound corresponds to 
min ${\rm \bar{P}_{MSW}(E)} \cong \sin^2\theta
\cong 8\times 10^{-3}$, and represents the 99.73\% (95\%) C.L. limit.
For values of  $\sin^2\theta$ different from the indicated one the suppression
due to ${\rm \bar{P}_{MSW}(E)}$ of the integral entering into the expression for
the contribution of $^{8}$B neutrinos in ${\rm \bar{R}(Ar)}$ (see, e.g., [11]),
as can be shown, is weaker and
the upper bound on $~{\rm f_{B}}$ one gets is not an absolute upper bound. 
Actually, $8\times 10^{-3}$ is the minimal value of $\sin^2\theta$ for which 
one has ${\rm \bar{P}_{MSW}(E)} \cong \sin^2\theta$ for all $^{8}$B neutrinos 
having energy ${\rm E \gtap 3.0~MeV}$. In the same way one can use the 
Kamiokande data to obtain an upper limit on $~{\rm f_{B}}$, which, however, 
turns out to be less constraining: for $\sin^2\theta \cong 8\times 10^{-3}$
one obtains $~{\rm f_{B}}\ltap 100~(85)$.  Obviously, 
the maximal value of $~{\rm f_{B}}$, allowed by the Cl--Ar data, will be 
smaller than that in eq. (8) (and may occur for a slightly different value of 
$\sin^2\theta$) if one takes into account the contributions due to the 
$^{7}$Be and the CNO neutrinos in ${\rm \bar{R}(Ar)}$, as will be done in our 
analysis. 

      The remarkable difference between the upper bounds on $~{\rm f_{B}}$
implied by the Kamiokande data in the cases of 
$\nu_{e} \rightarrow \nu_{\mu (\tau)}$ and of 
$\nu_{e} \rightarrow \nu_{s}$ MSW transitions has the following
origin. The maximal allowed values of $~{\rm f_{B}}$ correspond to maximal
possible suppression of ${\rm \bar{R}(K)}$ due to the MSW transitions, being 
inversely proportional to the relevant (effective) suppression factor 
(see, e.g., refs. [22,11]). Since the active
neutrinos $\nu_{\mu (\tau)}$ contribute to the Kamiokande signal, while the
sterile neutrinos $\nu_{s}$ do not, the signal due to the $^{8}$B neutrinos
having energy E is suppressed in the cases of 
$\nu_{e} \rightarrow \nu_{\mu (\tau)}$ and  $\nu_{e} \rightarrow \nu_{s}$ 
transitions respectively by the probability factors
$[{\rm \bar{P}_{MSW}(E)} + 0.15~(1 - {\rm \bar{P}_{MSW}(E)})]$ and 
${\rm \bar{P}_{MSW}(E)}$,
where the term with the coefficient 0.15 is due to the $\nu_{\mu (\tau)}$
contribution to ${\rm \bar{R}(K)}$ [22].
As a consequence of the specific dependence of ${\rm \bar{P}_{MSW}(E)}$ on
${\rm E/\Delta m^2}$ and $\sin^22\theta$ the strongest possible suppression of
${\rm \bar{R}(K)}$ due to the transitions 
$\nu_{e} \rightarrow \nu_{s}$ is by the factor $\sim (6 - 8)\times 10^{-3}$, 
and it is by a much bigger factor $\sim~$0.15 when the transitions 
are into active neutrino $\nu_{\mu (\tau)}$. 

  In the present study of the stability of the MSW 
$\nu_{e} \rightarrow \nu_{s}$ transition solution with respect 
to ${\rm \Phi_{B}}$ and ${\rm \Phi_{Be}}$ variations, we use the approach
adopted in the similar studies of the MSW 
$\nu_{e} \rightarrow \nu_{\mu (\tau)}$ transition [23] and of the 
vacuum oscillation
[22] solutions. Namely, the fluxes of the pp, pep and the CNO neutrinos 
(see, e.g., refs. [2])
are kept fixed and their values were taken from ref. [4]. The fluxes of the 
$^{8}$B and $^{7}$Be neutrinos, and
correspondingly, ${\rm f}_{{\rm B}}$ and f$_{{\rm Be}}$, are treated as fixed
parameters, which, however, are allowed to take any values within certain
intervals. In the case of ${\rm \Phi_{Be}}$ the interval chosen corresponds to
$$0.7 \leq {\rm f}_{{\rm Be}} \leq 1.3.~~~~\eqno(9)$$
\noindent It is somewhat wider than the interval formed by the current solar
model predictions: 0.86 -- 1.06. For ${\rm \Phi_{B}}$ values in the interval 
determined by the inequalities (6) and (8) were considered. 

    The above approach is
motivated by the fact that the contributions of the CNO neutrinos to the
signals in all three types of detectors [1,12--14] are predicted to be
relatively small [2,4--9], and that (apart from the CNO neutrinos) the spreads 
in the predictions for the fluxes ${\rm \Phi_{B}}$ and ${\rm \Phi_{Be}}$ are 
the largest. A more detailed discussion and further justification of this 
approach are given in refs. [23,22] (see also [11]). We would like only
to emphasize here that the aim of our study (as like of the studies of the MSW 
$\nu_{e} \rightarrow \nu_{\mu (\tau)}$ transition and of the vacuum oscillation
solutions performed in refs. [23,22]) was, in particular, to determine the 
ranges of values of ${\rm \Phi_{B}}$ and ${\rm \Phi_{Be}}$ for which the 
possibility of MSW $\nu_{e} \rightarrow \nu_{s}$ transitions of solar 
neutrinos cannot be excluded by the existing solar neutrino data. Certainly,
values of ${\rm \Phi_{B}}$ corresponding to, e.g., f$_{{\rm B}} \cong~$30
cannot be expected at present to appear in any realistic solar model.

%Some of the values of ${\rm \Phi_{Be}}$
%used in the analyses, as those corresponding to f$_{{\rm Be}}=~$0.7 and 1.3,
%for example, are incompatible with the constraint on the solar neutrino fluxes
%which the data on the solar luminosity impose (see, e.g., refs. [11,28]):
%$$\Phi_{\rm pp} + 0.958\Phi_{\rm Be} + 0.955 \Phi_{\rm CNO} + 0.910\Phi_{\rm
%pep}= (6.517~\pm~0.02) \times 10^{10}~{\rm cm}^{-2}{\rm sec}^{-1},
%~~\eqno (10)$$
%\noindent where $\Phi_{\rm CNO} = \Phi_{\rm N} + \Phi_{\rm O}$,
%and $\Phi_{\rm pp}$, $\Phi_{\rm pep}$, $\Phi_{\rm N}$ and $\Phi_{\rm O}$ are
%the fluxes of the pp, pep and the CNO neutrinos.
%However, a 20\% -- 30\% change in ${\rm \Phi_{Be}}$ with respect to
%${\rm \Phi_{Be}^{BP}} = 4.89 \times 10^{9}~\nu_e$/cm$^{2}$/sec
%is required by (9) to be balanced by only a few percent
%change of the pp neutrino flux, and the latter will have a small effect on
%the predictions for the signal in the Ga--Ge experiments [13,14]. Besides, the
%aim of our study 
%(as like of the studies of the MSW solutions performed in refs. [25,26]) 
%was, in particular, to determine the ranges of values of
%${\rm \Phi_{B}}$ and ${\rm \Phi_{Be}}$ for which the possibilities of MSW
%$\nu_{e} \rightarrow \nu_{s}$ transitions of solar neutrinos cannot 
%be excluded by the existing solar neutrino data. Certainly,
%values of ${\rm \Phi_{B}}$ corresponding to, e.g., f$_{{\rm B}} \cong~$30
%cannot be expected to appear in any realistic solar model.

   In the absence of ``unconventional'' behavior (MSW transitions, 
vacuum oscillations, etc.) of solar neutrinos,
the signals in the Cl--Ar and Ga--Ge experiments can be written in the
following form within the above approach:
$${\rm \bar{R}(Ar)} = (6.20{\rm f_{B}} + 1.17{\rm f_{Be}} + 0.40_{\rm CNO} +
0.23_{\rm pep})~{\rm SNU},~~\eqno (10)$$
$${\rm \bar{R}(Ge)} = (70.8_{\rm pp} + 3.1_{\rm pep} + 35.8{\rm f_{Be}} +
   13.8{\rm f_{B}} + 7.9_{\rm CNO})~{\rm SNU},~~\eqno (11)$$

\noindent where 6.20${\rm f_{B}~SNU}$ is the contribution in 
${\rm \bar{R}(Ar)}$ due to the $^{8}$B neutrinos, etc.

   We have applied the $\chi^2-$method in the present analysis. In computing
the $\chi^2$ for a given pair of values of the parameters $\Delta m^2$ and
$\sin^22\theta$ we have neglected the estimated uncertainties in the reference
model predictions [4] for the solar neutrino fluxes since the ranges within
which ${\rm \Phi_{B}}$ and ${\rm \Phi_{Be}}$ have been varied exceed by far the
uncertainties. We did, however, take into account the uncertainties in the
detection cross--sections for the detectors [1,12--14].
   
  A recent study [19] of the solar $\nu_{e} \rightarrow \nu_{s}$ MSW
transition hypothesis, performed within the reference model [4] 
(${\rm f_{B}} = {\rm f_{Be}} = 1$) including the estimated uncertainties in the
predictions for the pp, pep, $^{7}$Be, $^{8}$B and CNO neutrino fluxes, has
shown that at 95\% C.L. there exists a small mixing angle nonadiabatic 
solution with 
$$3.0\times 10^{-6}~{\rm eV}^2 \ltap \Delta m^2
     \ltap 8.1\times 10^{-6}~{\rm eV}^2,~~~~\eqno(12a)$$ 
$$3.3\times 10^{-3} \ltap \sin^22\theta \ltap 1.3\times 10^{-2},
~~~~\eqno(12b)$$
\noindent which provided a good quality of the fit of the data available by
April 1995 \footnote{In July 1995 the GALLEX collaboration updated their
results adding new data from 9 runs of measurements to the previously existing
data from 30 runs [13].} (min $\chi^2 =
1.65$ for 2 d.f.). A large mixing angle (adiabatic) solution (present 
in the case of $\nu_{e} \rightarrow \nu_{\mu (\tau)}$ MSW transitions [18,19])
was shown [19] to be excluded at 98\% C.L. by the data.

  Allowing ${\rm f_{B}}$ and ${\rm f_{Be}}$ to take values in the intervals
(6), (8) and (9), respectively, we have searched for 
$\nu_{e} \rightarrow \nu_{s}$ MSW transition solution in the region
$10^{-9}~{\rm eV}^2 \leq \Delta m^2 \leq 10^{-4}~{\rm eV}^2$ and  
$10^{-4} \leq\sin^22\theta \leq 1$. It was found that the values of 
${\rm f_{B}}$ for which the hypothesis of solar $\nu_{e} \rightarrow \nu_{s}$ 
transitions provides at 95\% C.L. a description of the solar neutrino data form
a wide interval, whose width depends somewhat on ${\rm f_{Be}}$. We give below
these 95\% C.L. allowed intervals for ${\rm f_{B}}$ in the cases of 
${\rm f_{Be}} = 0.7; 1.0; 1.3$:
$${\rm f}_{{\rm Be}} = 0.7,~~~0.39 \ltap {\rm f}_{{\rm B}} \ltap 40,
       ~~~~~~~~\eqno (13a)$$
$${\rm f}_{{\rm Be}} = 1.0,~~~
0.39 \ltap {\rm f}_{{\rm B}} \ltap 37,~~~~~~~~\eqno (13b)$$
$${\rm f}_{{\rm Be}} = 1.3,~~~
0.40 \ltap {\rm f}_{{\rm B}} \ltap 36.~~~~~~~~\eqno (13c)$$

 The maximal values of ${\rm f_{B}}$ in eqs. (13a) -- (13c) are determined by 
the
Cl--Ar data (1) and therefore exhibit a certain, although weak, dependence on 
${\rm f_{Be}}$. The value of ${\rm f_{B}}$ is constrained from below
basically by the Kamiokande result (2). The corresponding allowed regions of
values of the parameters $\Delta m^2$ and $\sin^22\theta$ lie all within the 
intervals:
$$3.0\times 10^{-6}~{\rm eV}^2 \ltap \Delta m^2
     \ltap 1.1\times 10^{-5}~{\rm eV}^2,~~~~\eqno(14a)$$ 
$$8.0\times 10^{-4} \ltap \sin^22\theta \ltap 0.45.~~~~\eqno(14b)$$

\noindent Note that the $\Delta m^2-$interval (14a) of all the solutions with 
$0.39 \ltap {\rm f}_{{\rm B}} \ltap 40$ is only slightly wider than the
interval (12a) of the values of $\Delta m^2$ of the ${\rm f_{B}} = 
{\rm f_{Be}} = 1$ solution. Our results are illustrated in Figs.
1a -- 1c where 95\% C.L. allowed regions in the $\Delta m^2 - \sin^22\theta$
plane for a number of values of ${\rm f_{B}}$ from the interval 
$0.40 \leq {\rm f}_{{\rm B}} \leq 35$ and for ${\rm f_{Be}} = 0.7; 1.0; 1.3$
are shown. The values of $\Delta m^2$ allowed for given 
${\rm f}_{{\rm B}}$ and ${\rm f}_{{\rm Be}}$ as well as the solution intervals
(12a) and (14a) are determined primarily by the Ga--Ge data, while the solution
values of $\sin^22\theta$ are constrained by the Kamiokande and the Cl--Ar
results.  

  For any given allowed by the data value of ${\rm f}_{{\rm B}} < 3$ the MSW 
$\nu_{e} \rightarrow \nu_{s}$ transition solution is of the nonadiabatic (NA) 
type, i.e., it corresponds to nonadiabatic transitions of at least a large
fraction (the higher energy) $^{8}$B neutrinos, while the transitions of the
pp, $^{7}$Be, pep and CNO neutrinos are adiabatic as long as 
$\sin^22\theta \gtap 4\times 10^{-3}$; for smaller values of $\sin^22\theta$
some of the pp and CNO neutrinos and the $^{7}$Be and pep neutrinos can 
undergo nonadiabatic transitions as well.
This solution is the $\nu_{e} \rightarrow \nu_{s}$ transition analog of the
MSW $\nu_{e} \rightarrow \nu_{\mu (\tau)}$ transition NA solution at 
${\rm f}_{{\rm B}} < 3$ discussed in detail in ref. [23]. The existence of the
NA solution of interest, the location of the allowed region corresponding to a
given value of ${\rm f}_{{\rm B}}$, the movement and the change of the size of
this region when the value of ${\rm f}_{{\rm B}}$ is changed, etc. can be
explained in the same way as this was done for the case of the
$\nu_{e} \rightarrow \nu_{\mu (\tau)}$ transition NA solution in ref. [23].

  As ${\rm f}_{{\rm B}}$ increases beyond the value $\sim~$2 the ``dynamics'' 
of the $\nu_{e} \rightarrow \nu_{s}$ transition solution is the following. At 
${\rm f}_{{\rm B}} \gtap 3$ an adiabatic (AD) solution appears in addition to
the nonadiabatic one for $3.0\times 10^{-2} \ltap \sin^22\theta \ltap 0.46$ 
and for
values of $\Delta m^2$ which lie within the interval (14a) (see Figs. 1a --
1c). For $3 \ltap {\rm f}_{{\rm B}} \ltap 20$ the regions of the NA and of the 
AD solutions in the $\Delta m^2 - \sin^22\theta$ plane are disconnected. With 
the  increase of ${\rm f}_{{\rm B}}$ the two
regions approach each other. They become continuously connected at 
${\rm f}_{{\rm B}} \sim 22$. As ${\rm f}_{{\rm B}}$ increases further
the subregion of the NA solution diminishes.
%and, e.g., for ${\rm f}_{{\rm Be}} = 0.7$ it practically disappears when 
%${\rm f}_{{\rm B}} \cong 35$.

  In contrast to the respective NA solutions at ${\rm f}_{{\rm B}} < 3$, 
the $\nu_{e} \rightarrow \nu_{\mu (\tau)}$ and the 
$\nu_{e} \rightarrow \nu_{s}$   
adiabatic solutions differ considerably: the first occurs for [23] (see also
the first article quoted in [18])
$1.0 \ltap {\rm f}_{{\rm B}} \ltap 3.4$, $0.15 \ltap \sin^22\theta \ltap 0.90$ 
and $6.2\times 10^{-6}~{\rm eV}^2 \ltap \Delta m^2 \ltap 10^{-4}~{\rm eV}^2$,
while the second holds for ${\rm f}_{{\rm B}} \gtap 3$ and typically 
smaller 
values of $\sin^22\theta$ and $\Delta m^2$, as Figs. 1a -- 1c indicate.
None of the allowed regions in the $\Delta m^2 - \sin^22\theta$ plane
corresponding to the two adiabatic solutions overlap (compare our Fig. 1a (1b) 
with Fig. 1b (1a) in ref. [23]).

  For given ${\rm f}_{{\rm B}} < 3$ and ${\rm f}_{{\rm Be}}$   
from (9) the region of the NA $\nu_{e} \rightarrow \nu_s$ solution is shifted
to smaller values of $\Delta m^2$ (on average by a factor of 1.2) with respect
to the region of the NA $\nu_{e} \rightarrow \nu_{\mu (\tau)}$ solution
corresponding to the same ${\rm f}_{{\rm B}}$ and ${\rm f}_{{\rm Be}}$ values
(compare Figs. 6a (6c) with Fig. 6b (6d) in ref. [19] as well as our Fig. 1a 
(1b) with Fig. 1b (1a) in ref. [23]). This difference is due to the fact that in
matter $\nu_{\mu (\tau)}$ scatters (coherently) on electrons, neutrons and
protons while $\nu _s$ does not. As a consequence the probabilities 
${\rm \bar{P}^{s}_{MSW}(E)}$ and ${\rm \bar{P}^{a}_{MSW}(E)}$ associated with 
the $\nu_{e} \rightarrow \nu_s$ and $\nu_{e} \rightarrow \nu_{\mu (\tau)}$
transitions depend on different density ratios [27], i.e., on 
${\rm (N^{0}_{e} - 
N^{0}_{n}/2)/N^{res}}$ and ${\rm N^{0}_{e}/N^{res}}$ respectively, where
${\rm N^{0}_{e}}$ and ${\rm N^{0}_{n}}$ are the electron and neutron number
densities in the point of $\nu_e$ production in the Sun, and 
${\rm N^{res}} = \Delta m^2 \cos2\theta/{\rm (2E \sqrt {2} G_{F})}$ is the
resonance density. One has in the region of $\nu_e$ production in the Sun
[4--9]: ${\rm N^{0}_{n}/N^{0}_{e}} \cong (0.45 - 0.20)$. The difference between
${\rm \bar{P}^{s}_{MSW}(E)}$ and ${\rm \bar{P}^{a}_{MSW}(E)}$ due to the 
finite value of
the ratio ${\rm N^{0}_{n}/N^{0}_{e}}$ is insignificant when
${\rm (N^{0}_{e} - N^{0}_{n}/2)/N^{res}} >> 1, \tan^22\theta$, or if 
${\rm N^{0}_{e}/N^{res}} << 1$. However, for the transitions of the 
$^{7}$Be, pp, pep and CNO neutrinos neither of the above conditions is
fulfilled, the difference between the two averaged MSW probabilities is not 
negligible and leads to the shift in $\Delta m^2$ between the 
$\nu_{e} \rightarrow \nu_s$ and $\nu_{e} \rightarrow \nu_{\mu (\tau)}$ 
allowed regions. 
 
   As we have mentioned earlier, for fixed ${\rm f}_{{\rm B}}$ and 
${\rm f}_{{\rm Be}}$ the solution values of $\sin^22\theta$ are determined
primarily by the Cl--Ar and the Kamiokande data. In the case of the adiabatic
$\nu_{e} \rightarrow \nu_s$ solution they can be easily understood
qualitatively due to the fact that the $^{8}$B neutrino flux is suppressed 
approximately by the (energy--independent) factor $\sin^2\theta$.
As a consequence, the allowed values of $\sin^2\theta$ for a given 
${\rm f}_{{\rm B}}$, corresponding to an adiabatic solution, are 
constrained by the Kamiokande and the Cl--Ar data to lie in the interval
(95\% C.L.):
$$0.37 \ltap {\rm f}_{{\rm B}}\sin^2\theta < 0.49,~~~\eqno(15)$$
\noindent where the upper limit follows from (1). For ${\rm f}_{{\rm B}} = 3$, 
for instance, eq. (15) implies $0.43 \ltap \sin^22\theta < 0.55$. The 
exact calculations for ${\rm f}_{{\rm Be}} = 0.7~(1.0)$ give in this case: 
$0.41 \ltap \sin^22\theta \ltap 0.46~(0.45)$ (see Figs. 1a -- 1c). The maximal 
allowed value of $\sin^22\theta$ is smaller than the one implied by eq. (15) 
because, in particular, the upper limit in (15) was obtained without taking 
into account the contributions of $^{7}$Be and (pep + CNO) neutrinos 
in ${\rm \bar{R}(Ar)}$. The upper bound on $\sin^22\theta$ following from (15)
is closer to the maximal solution value of $\sin^22\theta$ for larger values 
of ${\rm f}_{{\rm B}}$, for which the indicated contributions in 
${\rm \bar{R}(Ar)}$ are smaller. In the case of, e.g., ${\rm f}_{{\rm B}} =
10~(20)$ from (15) one finds $\sin^22\theta < 0.19~(0.096)$, while our results
(Figs. 1a -- 1c) show that $\sin^22\theta \ltap 0.18~(0.094)$.

   If $0.40 \ltap {\rm f}_{{\rm B}} \leq 1$ the hypothesis of MSW 
$\nu_{e} \rightarrow \nu_{s}$ transitions of solar neutrinos provides a 
good quality of the fit of the data (1) -- (4): for ${\rm f}_{{\rm Be}} =
0.7;~1.3$ we have ${\rm min}~\chi^2 = 1.22;~1.27$ (for 2 d.f.) reached at 
${\rm f}_{{\rm B}} = 1.0;~0.99$, 
$\Delta m^2 \cong 4.3\times 10^{-6}~{\rm eV^2}$ 
and $\sin^22\theta \cong 7.6\times 10^{-3}$.
A somewhat better description of the data is achieved for larger values of 
${\rm f}_{{\rm B}}$: if we fix ${\rm f}_{{\rm Be}} = 1.0~(0.7)$ and consider 
the interval $0.40 \leq {\rm f}_{{\rm B}} \leq 3.5$, the minimal value of the 
$\chi^2-$function occurs at ${\rm f}_{{\rm B}} \cong 3.4$,
$\Delta m^2 \cong 4.3~(4.6)\times 10^{-6}~{\rm eV^2}$ and
$\sin^22\theta \cong 1.9~(1.8)\times 10^{-2}$, and one has
${\rm min}~\chi^2 = 0.85~(0.82)$. For the adiabatic solution with 
${\rm f}_{{\rm B}} \leq 5$ and, e.g., ${\rm f}_{{\rm Be}} = 1.0$, one finds 
${\rm min}~\chi^2 = 4.3$ at ${\rm f}_{{\rm B}} \cong 5.0$, 
$\Delta m^2 \cong 6.3\times 10^{-6}~{\rm eV^2}$ and $\sin^22\theta \cong 0.29$.
\vskip 0.3cm
\leftline {\bf 3. PHYSICAL IMPLICATIONS OF THE MSW 
$\nu_{e} \rightarrow \nu_{s}$ SOLUTIONS}
\vskip 0.3cm
     We shall discuss next the physical implications of the MSW
$\nu_{e} \rightarrow \nu_{s}$ transition solutions for the future solar
neutrino experiments SNO [28], Super--Kamiokande [29], BOREXINO [30] and 
HELLAZ [31]. Consider first the ``low'' $^{8}$B neutrino flux solution, 
$0.4 \ltap {\rm f}_{{\rm B}} \ltap 0.6$, which holds for small values of 
$\sin^22\theta$: $8.0\times 10^{-4} \ltap \sin^22\theta \ltap 
4.5\times 10^{-3}$. For this solution the $^{8}$B $\nu_e$ spectrum and the 
corresponding spectrum of the recoil $e^{-}$ from the reaction 
$\nu + e^{-} \rightarrow \nu + e^{-}$ due to $^{8}$B neutrinos to be measured
respectively in SNO and Super--Kamiokande experiments are predicted to be
relatively weakly deformed (see Figs. 2a and 2b). The deformation of the 
$^{8}$B $\nu_e$ spectrum 
caused for given $\Delta m^2$ and $\sin^22\theta$ by the MSW 
$\nu_{e} \rightarrow \nu_{s}$ transitions practically 
coincides with the deformation
produced by the MSW $\nu_{e} \rightarrow \nu_{\mu (\tau)}$ transitions 
for the same values of the two parameters. The deformations of the 
recoil $e^{-}$ spectrum generated by the two types of MSW transitions of 
$^{8}$B neutrinos differ in general. This difference can be quite substantial
for relatively large values of $\sin^22\theta$, but diminishes with the 
decreasing of $\sin^22\theta$ and becomes hardly observable for 
$\sin^22\theta \ltap 4\times 10^{-3}$, i.e., for the ``low'' 
$\Phi_{{\rm B}}$ solution under discussion (compare our Figs.
2a and 2b respectively with Figs. 4c, 4d and 6c, 6d in ref. [19] and with 
Figs. 2 and 3 in ref. [23]).
The ratio ${\rm R_{CC/NC}}$ of the rates of events
due to the charged current (CC) and neutral current (NC) reactions
$\nu_e + D \rightarrow e^{-} + p + p$ and $\nu + D \rightarrow \nu + p + n$
by which the $^{8}$B neutrinos (with energy ${\rm E \geq 6.44~MeV}$ and  
${\rm E \geq 2.2~MeV}$, respectively) will be detected in the SNO experiment,
is not sensitive to the transitions (or oscillations) into sterile neutrino
\footnote{For the ``low'' $\Phi_{{\rm B}}$ MSW 
$\nu_{e} \rightarrow \nu_{\mu (\tau)}$ solution on has [23]: 
${\rm R_{CC/NC}} \cong 0.75~(0.85){\rm R^{SM}_{CC/NC}}$ for ${\rm f}_{{\rm B}}
= 0.5~(0.4)$, where 
${\rm R^{SM}_{CC/NC}}$ is the ratio of the CC and NC event rates predicted in
the absence of ``unconventional'' behavior of the $^{8}$B neutrinos.}.

   Possible signature of the ``low'' $\Phi_{{\rm B}}$ solution under discussion
could be i) a sufficiently strong suppression of the flux of 0.862 MeV 
$^{7}$Be neutrinos by a factor not exceeding, e.g., 0.15, or ii) a specific
deformation of the spectrum of pp neutrinos having energy ${\rm E \gtap
0.22~MeV}$. The 0.862 MeV $^{7}$Be neutrinos are predicted to produce
approximately 90\% of the signal in the BOREXINO detector [30], while the 
HELLAZ detector [31] is envisaged to measure the total flux and the spectrum 
of pp neutrinos with ${\rm E \geq 0.218~MeV}$. In both detectors the 
$\nu - e^{-}$ elastic scattering reaction will be utilized for detection 
of solar neutrinos. The pp $\nu_e$ spectrum will be reconstructed from
the data the HELLAZ detector is conceived to provide about the 
recoil-electron spectrum in the e$^{-}$ kinetic energy region 
${\rm T_{e} \geq 0.1~MeV}$, and the measurement of the recoil-electron 
momentum direction \footnote{From the kinematics of the reaction
$\nu + e^{-} \rightarrow \nu + e^{-}$ it follows that for the initial neutrino
(recoil-e$^{-}$ kinetic) energy ${\rm E \leq 0.42~MeV}$ 
(${\rm T_{e} \geq 0.10~MeV}$) one has ${\rm T_{e} \leq 0.26~MeV}$
(${\rm E \geq 0.218~MeV}$). This determines the energy intervals in which 
the two spectra are expected to be measured with HELLAZ: 
${\rm 0.218~MeV \ltap E \ltap 0.42~MeV}$ and 
${\rm 0.10~MeV \ltap T_{e} \ltap 0.26~MeV}$.}.   
   
   Since the BOREXINO detector is based on the reaction 
$\nu + e^{-} \rightarrow \nu + e^{-}$ [30], the active neutrinos 
$\nu_{\mu (\tau)}$ are going to contribute
to the BOREXINO signal, while the sterile neutrinos $\nu_s$ will not. As a
consequence, the event rate generated by the 0.862 MeV $^{7}$Be neutrinos in
BOREXINO can be suppressed at most by the factor 0.21 if they undergo 
$\nu_{e} \rightarrow \nu_{\mu (\tau)}$ transitions, and it can be reduced by a
noticeably smaller factor when the transitions are of the 
$\nu_{e} \rightarrow \nu_{s}$ type. In the latter case the suppression factor
coincides with the averaged MSW probability ${\rm \bar{P}^{s~Be}_{MSW}} =
{\rm \bar{P}^{s}_{MSW}(E=0.862~MeV)}$.

   If $0.4 \ltap {\rm f}_{{\rm B}} \ltap 0.6$ and we consider only the
corresponding values of $\Delta m^2$ and $\sin^22\theta$ allowed (at 95\% C.L.)
by the data (1) -- (4), ${\rm \bar{P}^{s~Be}_{MSW}}$ is constrained to lie in
the interval $\sim (0.06 - 0.40)$, the minimal and maximal values of
${\rm \bar{P}^{s~Be}_{MSW}}$ taking place at $\Delta m^2 \cong 
3.1\times 10^{-6}~{\rm eV^2}$  and $\Delta m^2 \cong 
8.0\times 10^{-6}~{\rm eV^2}$, but at the same $\sin^22\theta \cong
4.5\times 10^{-3}$. One has ${\rm \bar{P}^{s~Be}_{MSW}} \leq 0.15$ for 
$5.3\times 10^{-6}~{\rm eV^2} \ltap \Delta m^2 \ltap 
8.0\times 10^{-6}~{\rm eV^2}$. The lowest (highest) value of this 
$\Delta m^2-$interval 
depends on the value of $\sin^22\theta$: the one given corresponds to 
$\sin^22\theta = 10^{-3}$ and it changes from 
$5.3~(8.0)\times 10^{-6}~{\rm eV^2}$ to $3.1~(7.4)\times 10^{-6}~{\rm eV^2}$ 
when $\sin^22\theta$ increases from $10^{-3}$ to $4.5\times 10^{-3}$. For 
values of $\Delta m^2$ in the solution interval 
$3.0\times 10^{-6}~{\rm eV^2} \ltap \Delta m^2 \ltap 
4.5\times 10^{-6}~{\rm eV^2}$ the MSW $\nu_{e} \rightarrow \nu_{s}$ transitions
lead to a relatively large distortion of the spectrum of pp ($\nu_e$) 
neutrinos (recoil electrons) having energy ${\rm E \gtap 0.30~MeV}$ 
\footnote{For the same values of 
$\Delta m^2$ and $\sin^22\theta$ the pp neutrino spectrum will be
deformed, but much weaker, by MSW $\nu_{e} \rightarrow \nu_{\mu
(\tau)}$ transitions.}  (${\rm T_{e} \gtap 0.16~MeV}$), which might be
observable with the HELLAZ detector; for $3.0\times 10^{-6}~{\rm eV^2} \ltap 
\Delta m^2 \ltap 3.7~(4.0)\times 10^{-6}~{\rm eV^2}$ the pp $\nu_e$ 
(recoil-electron) spectrum will be strongly deformed in the entire neutrino
(electron) energy interval ${\rm E = (0.218 - 0.42)~MeV}$ 
(${\rm T_{e} = (0.10 - 0.26)~MeV}$) relevant to the HELLAZ detector
\footnote{A suppression of the signal in BOREXINO due to the 
0.862 MeV $^{7}$Be neutrinos by a factor not exceeding 0.15 concomitant with 
a strong distortion of the pp $\nu_e$ spectrum is predicted in the case of the 
``low'' $\Phi_{{\rm B}}$, $0.35 \ltap {\rm f}_{{\rm B}} \ltap 0.43$, vacuum
$\nu_e \leftrightarrow \nu_s$ oscillation solution of the solar neutrino
problem [22]. However, the $\nu_e \leftrightarrow \nu_s$ oscillations would 
deform the lower energy part of the pp $\nu_e$ spectrum, ${\rm E \ltap
0.30~MeV}$, and the distortions will be very different from those caused by 
the $\nu_e \rightarrow \nu_s$ MSW transitions (compare our Fig. 3a with 
Fig. 3b in ref. [22]).}. This is 
illustrated in Fig. 3a and Figs. 4a -- 4c, where the pp $\nu_e$ spectrum 
and the spectrum of the recoil electrons from the reaction 
$\nu + e^{-} \rightarrow \nu + e^{-}$ induced by the pp neutrinos,
deformed by the MSW $\nu_{e} \rightarrow \nu_{s}$ transitions under discussion,
are shown for $\Delta m^2 = 3.0; 3.5; 4.0; 4.5; 5.0~\times 10^{-6}~{\rm eV^2}$
and respectively for $\sin^22\theta = 5\times 10^{-3}$ and 
$\sin^22\theta = 10^{-3},~5\times 10^{-3},~10^{-2},~2\times 10^{-2}$. As our 
calculations and Figs. 4a -- 4c show, for a given value
of $\Delta m^2$ from the interval $3.0~\times 10^{-6}~{\rm eV^2}\ltap 
\Delta m^2 \ltap 5.0~\times 10^{-6}~{\rm eV^2}$ of interest the corresponding
deformed pp $\nu_e$ and recoil-electron spectra change very little with the 
change of $\sin^22\theta$ within the interval $10^{-3} \ltap \sin^22\theta 
\ltap 2\times 10^{-2}$. Therefore Fig. 3a actually illustrates the pp 
$\nu_e$ spectrum deformations which are caused by the MSW 
$\nu_{e} \rightarrow \nu_{s}$ transitions when ${\rm f}_{{\rm B}}$ has a value
in the wider interval $0.4 \ltap {\rm f}_{{\rm B}} \ltap 3.0$.
If $\Delta m^2 \gtap 6\times 10^{-6}~{\rm eV^2}$ 
the pp neutrino flux suppression 
factor reads ${\rm \bar{P}^{s~pp}_{MSW}(E)} \cong 1 - 1/2 \sin^22\theta$, 
${\rm E \leq 0.42~ MeV}$, and its deviation from 1 is unobservable. 

   The $\nu_{e} \rightarrow \nu_{s}$ NA solution at $0.6 < {\rm
f}_{{\rm B}} \ltap 2.0$, for which $5\times 10^{-3}\ltap \sin^22\theta
\ltap 2\times 10^{-2}$ and $3.3\times 10^{-6}~{\rm eV^2} \ltap \Delta
m^2 \ltap 8.0\times 10^{-6}~{\rm eV^2}$, implies (for given $\Delta
m^2$ and $\sin^22\theta$) the same distinctive deformation of the
$^{8}$B $\nu_e$ spectrum as the analogous $\nu_{e} \rightarrow
\nu_{\mu (\tau)}$ NA solution.  Detailed studies of the $^{8}$B
$\nu_e$ spectrum deformation due to the MSW transitions in the case of
interest can be found in refs. [19,23,32] (see also Fig. 2a). However,
the distortion of the spectrum of the recoil electron from the
reaction $\nu + e^{-} \rightarrow
\nu + e^{-}$ caused by $^{8}$B neutrinos to be measured by the
Super--Kamiokande collaboration is predicted to be larger than that
implied by the similar $\nu_{e} \rightarrow \nu_{\mu (\tau)}$ solution
for the same $\Delta m^2$ and $\sin^22\theta$ (compare our Fig. 2b
with Figs. 6c, 6d in ref. [19] and with Fig. 3 in ref. [23]). By
measuring the two spectra (in SNO and Super--Kamiokande experiments)
and by testing the different types of correlations between them the
$\nu_{e} \rightarrow \nu_{s}$ and the $\nu_{e} \rightarrow \nu_{\mu
(\tau)}$ transitions lead to, it can be possible to distinguish
between the MSW nonadiabatic $\nu_{e} \rightarrow \nu_{s}$ and
$\nu_{e} \rightarrow \nu_{\mu (\tau)}$ solutions for $0.6 < {\rm
f}_{{\rm B}} \ltap 2.0$. Other possible tests of the $^{8}$B neutrino
$\nu_{e} \rightarrow \nu_{s}$ conversion hypothesis, which could be
performed using the SNO and Super--Kamiokande data provided the
relevant conversion probability exhibits a nontrivial neutrino energy
dependence (as in the case of the NA solution under discussion), were
proposed in ref. [33].  Further, if $0.6 < {\rm f}_{{\rm B}} \ltap
2.0$, the 0.862 MeV $^{7}$Be $\nu_e$ flux is predicted to be
suppressed due to the NA $\nu_{e} \rightarrow \nu_{s}$ transitions by
a factor ${\rm \bar{P}^{s~Be}_{MSW}} \leq 0.15$ for most of the values
of $\Delta m^2$ and $\sin^22\theta$ from the allowed region
corresponding to given ${\rm f}_{{\rm B}}$ and ${\rm f}_{{\rm
Be}}$. For instance, for $\sin^22\theta = 5\times
10^{-3};~10^{-2};~2\times 10^{-2}$ we have ${\rm \bar{P}^{s~Be}_{MSW}}
\leq 0.15$ if respectively $\Delta m^2 \ltap 7.3;~7.0;~6.8~\times
10^{-6}~{\rm eV^2}$.  For the solution under discussion ${\rm
\bar{P}^{s~Be}_{MSW}}$ takes values in the interval $\sim (0.003 -
0.40)$, the maximal value being reached for $\Delta m^2 \cong
8.0\times 10^{-6}~{\rm eV^2}$ and $\sin^22\theta \cong 5\times
10^{-3}$.  If $\Delta m^2 > 5.0\times 10^{-6}~{\rm eV^2}$ the pp
$\nu_e$ flux will be practically unaffected by the NA $\nu_{e}
\rightarrow \nu_{s}$ transitions as ${\rm \bar{P}^{s~pp}_{MSW}(E)}
\cong 1 - 1/2 \sin^22\theta$, while for smaller values of
$\Delta m^2$ they will cause a rather large (and, perhaps, observable)
distortion of the pp $\nu_e$ and the corresponding recoil-electron
spectra (Figs. 3a, 4a -- 4d).

{} For ``large'' values of $\Phi_{{\rm B}}$, ${\rm f}_{{\rm B}} >
2.0$, there exists a $\nu_{e} \rightarrow \nu_{s}$ NA solution, but
not a similar $\nu_{e} \rightarrow \nu_{\mu (\tau)}$ one. The $\nu_{e}
\rightarrow \nu_{s}$ solution holds for $10^{-2} \ltap \sin^22\theta
\ltap 6\times 10^{-2}$ (Figs. 1a -- 1c). Its distinctive features are:
i) characteristic distortions of the $^{8}$B $\nu_e$ (${\rm E \geq
6.44~MeV}$) and of the corresponding recoil--electron (${\rm E_{e}
\geq 5.0~MeV}$) spectra (Figs. 2a and 2b), ii) ${\rm R_{CC/NC}} \cong
{\rm R^{SM}_{CC/NC}}$, iii) ${\rm \bar{P}^{s~Be}_{MSW}} \leq 0.15$
occurring for, e.g., $\sin^22\theta = 10^{-2};~2\times
10^{-2};~4\times 10^{-2}$ in the region $\Delta m^2 \ltap
7.0;~6.8;~6.7 ~\times 10^{-6}~{\rm eV^2}$, and iv) substantial
distortion of the pp $\nu_e$ (and the recoil-e$^{-}$) spectrum for
$\Delta m^2 \ltap 4.5\times 10^{-6}~{\rm eV^2}$. Actually, with the
change of ${\rm f}_{{\rm B}}$, $\Delta m^2$ and $\sin^22\theta$ within
their respective intervals the 0.862 MeV $^{7}$Be neutrino suppression
factor ${\rm \bar{P}^{s~Be}_{MSW}}$ is predicted to change from $\sim
3\times 10^{-3}$ (at $\Delta m^2 \cong 3.3\times 10^{-6}~{\rm eV^2}$)
to $\sim 0.34$ (for $\Delta m^2 \cong 8.0\times 10^{-6}~{\rm
eV^2}$). The minimal value of ${\rm \bar{P}^{s~Be}_{MSW}}$ depends
strongly on $\sin^22\theta$ increasing from $3\times 10^{-3}$ to
approximately $10^{-2}~(6\times 10^{-3})$ when $\sin^22\theta$ changes
from $10^{-2}$ to $4\times 10^{-2}~(2\times 10^{-2})$.  At the same
time the maximal value varies insignificantly (decreasing by $\sim
0.02$) with the variation of $\sin^22\theta$.

{} For the $\nu_{e} \rightarrow \nu_{s}$ AD solution, which we have
found to exist at ${\rm f}_{{\rm B}} \gtap 3.0$ (which is strongly
incompatible with the current solar model predictions [4--9] and which
we shall consider briefly for completeness), the $^{8}$B $\nu_e$ flux
will be suppressed by the energy independent factor ${\rm
\bar{P}^{s~}_{MSW}}
\cong \sin^2\theta$ and there will be no distortions of the $^{8}$B 
$\nu_e$ and of the corresponding recoil-electron spectra. One will have also 
${\rm R_{CC/NC}} \cong {\rm R^{SM}_{CC/NC}}$. In contrast, in the case of the
$\nu_{e} \rightarrow \nu_{\mu (\tau)}$ AD solution possible if
$1.0 \ltap {\rm f}_{{\rm B}} \ltap 3.4$ [23], one has [19] 
${\rm R_{CC/NC}} \leq 0.4~{\rm R^{SM}_{CC/NC}}$ 
\footnote{For values
of $\Delta m^2 \ltap 2.0\times 10^{-5}~{\rm eV^2}$ from the region of the 
$\nu_{e} \rightarrow \nu_{\mu (\tau)}$ AD solution we have: 
${\rm R_{CC/NC}} \cong \sin^2\theta~{\rm R^{SM}_{CC/NC}}$, with
$0.06 \ltap \sin^2\theta \ltap 0.34$.},
and the spectra of the $^{8}$B $\nu_e$ having ${\rm E \geq 6.44~MeV}$ and of
the recoil e$^{-}$ with  ${\rm E_{e} \geq 5.0~MeV}$ can be strongly deformed
(see the first article quoted in ref. [18] and ref. [19]). The indicated strong
spectra deformations are predicted to take place in the $\nu_e$ and e$^{-}$
energy intervals (6.44 -- 11.0) MeV and (5.0 -- 11.0) MeV, respectively, for 
$\Delta m^2 \gtap (4.5 - 5.0)\times 10^{-5}~{\rm eV^2}$ and are clearly 
distinguishable [19] from, e.g., those associated with the 
$\nu_{e} \rightarrow \nu_{\mu (\tau)}$ NA solution at $0.6 < {\rm f_{B}} 
\ltap 2.0$ (see Figs. 4c, 5c and 6c in ref. [19]). For the 
$\nu_{e} \rightarrow \nu_{s}$ AD solution the allowed values of
$\Delta m^2$ and $\sin^22\theta$ for, e.g., $3.0 \ltap {\rm f}_{{\rm
B}} \ltap 5.0$ lie in the region $5.0\times 10^{-6}~{\rm eV^2} \ltap
\Delta m^2 \ltap 8.6 \times 10^{-6}~{\rm eV^2}$, $0.26 \ltap
\sin^22\theta \ltap 0.44$. In this case we have $0.15 \ltap {\rm
\bar{P}^{s~Be}_{MSW}} \ltap 0.37$. The pp $\nu_e$ flux will be
suppressed by a factor (0.7 - 0.8), and the pp $\nu_e$ and
the associated recoil-electron spectra will be deformed for ${\rm E
\geq 0.22~MeV}$ and ${\rm T_{e} \gtap 0.1~MeV}$ (Figs. 3b and 5). Note
that since the total pp neutrino flux $\Phi_{{\rm pp}}$ is tightly
constrained by the data on the solar luminosity (see, e.g., [11]), an
observation of a value of $\Phi_{{\rm pp}} \leq 0.85 \Phi_{{\rm
pp}}^{{\rm SM}}$, where $\Phi_{{\rm pp}}^{{\rm SM}}$ is the pp
neutrino flux predicted, e.g., in ref.  [4], will be a very strong
indication for an ``unconventional'' behavior of the solar electron
neutrinos on their way to the Earth.

   There exist cosmological constraints on the MSW 
$\nu_{e} \rightarrow \nu_{s}$ transitions [34] (see also [35]), which are 
relevant to the solution of the solar neutrino problem under discussion. 
They follow from the upper limit on the (effective) number of neutrino species,
${\rm N}_{\nu}^{{\rm eff}}$, implied by big bang nucleosynthesis and helium 
abundance considerations [36]. A limit of ${\rm N}_{\nu}^{{\rm eff}} \leq 
3.1;~3.4;~3.8~$ is incompatible with the MSW adiabatic 
$\nu_{e} \rightarrow \nu_{s}$ transition solution for 
${\rm f}_{{\rm B}} \ltap 30;~9;~3$. The cosmological upper limit on 
${\rm N}_{\nu}^{{\rm eff}}$ is presently a subject of 
debate [36] with values ranging from less than 3 to approximately 4. 
\vskip 0.3cm   

\leftline {\bf 4. CONCLUSIONS}
\vskip 0.3cm
    The MSW $\nu_{e} \rightarrow \nu_{s}$ transition solution of the
solar neutrino problem is stable with respect to changes in the
predictions for the $^{8}$B and $^{7}$Be neutrino fluxes: for $0.7
\ltap {\rm f}_{{\rm Be}} \ltap 1.3$ the data (1) -- (4) admits
nonadiabatic and adiabatic $\nu_{e} \rightarrow \nu_{s}$ transition
solution for $0.4 \ltap {\rm f}_{{\rm B}} \ltap 36$ and $3 \ltap {\rm
f}_{{\rm B}} \ltap 36$ respectively. For $0.6 < {\rm f}_{{\rm B}}
\ltap 36$ the NA solution implies strong distortions of both the
$^{8}$B $\nu_e$ spectrum, measurable, e.g. in the CC disintegration of
the deuteron in SNO, and of the spectrum of the recoil electron from
the reaction $\nu + e^{-} \rightarrow \nu + e^{-}$ induced by $^{8}$B
neutrinos, measurable both in SNO and Super--Kamiokande experiments.
This solution can be distinguished from the $\nu_{e} \rightarrow
\nu_{\mu (\tau)}$ NA solution at $0.6 < {\rm f}_{{\rm B}} \ltap 2$ (which 
implies weaker distortion of the spectrum of recoil electrons) by
comparing the two spectra. Thus, if these spectra will be measured
with the planned accuracy in the two experiments, they can be used not
only to test the two solutions, but also to discriminate between them.

Additional possible signatures of the $\nu_{e} \rightarrow \nu_{s}$ NA
solution under discussion are: i) a strong suppression of the signal
due to the 0.862 MeV $^{7}$Be neutrinos in the BOREXINO detector by a
factor ${\rm \bar{P}^{s~Be}_{MSW}} \ltap 0.15$ (not possible in the
case of, e.g., $\nu_{e} \rightarrow\nu_{\mu (\tau)}$ MSW transitions),
and/or ii) a specific and rather strong distortion (typically of the
higher energy part, ${\rm E \gtap 0.30~MeV}$) of the pp $\nu_e$
spectrum at ${\rm E \gtap 0.218~MeV}$ and of the corresponding
recoil-electron spectrum at $0.10~{\rm MeV\ltap T_{e} \ltap
0.26~MeV}$. If $0.4 \ltap {\rm f}_{{\rm B}} \ltap 0.6$ these two
signatures can be the only ones by which the $\nu_{e} \rightarrow
\nu_{s}$ solution can be distinguished from the analogous $\nu_{e}
\rightarrow \nu_{\mu (\tau)}$ MSW solution as in both cases the
distortions of the $^{8}$B $\nu_e$ and the recoil-electron (${\rm
T_{e}\gtap 5.0~MeV}$) spectra can be unobservably small. Finally, the
$\nu_{e} \rightarrow \nu_{s}$ AD solution for $3.0 \ltap {\rm f}_{{\rm
B}} \ltap 5.0$ (which is a totally unrealistic possibility from the
point of view of the solar models currently discussed in the
literature) implies: i) an energy-independent reduction of the $^{8}$B
$\nu_e$ flux by a factor $0.07 \ltap \sin^2\theta \ltap 0.13$, ii) a
suppression of the BOREXINO signal due to the $^{7}$Be neutrinos by a
factor (0.33 - 0.50), iii) a relatively strong suppression (by
a factor (0.7 - 0.8) of the total pp $\nu_e$ flux, and iv) a
characteristic but not strong distortion of the pp $\nu_e$
(recoil-electron) spectrum (Figs. 3b and 5).  The last two signatures
can be crucial for the identification of the $\nu_{e} \rightarrow
\nu_{s}$ AD solution.
\vskip 0.3cm
\leftline{\bf Acknowledgments.} The work of P.I.K. was partially 
supported by funds from the Institute for Advanced
Study. S.T.P. wishes to thank O. P\`ene and the other colleagues from
L.P.T.H.E., Centre d'Orsay, where part of the work for this study has
been done, for the kind hospitality extended to him during his
visit. The work of S.T.P. was supported in part by the EEC grant
ERBCHRX CT930132.

\newpage
\centerline{\bf Figure Captions}
\medskip
\noindent
{\bf Figs. 1a -- 1c.} Regions of values of the parameters
$\Delta m^2$ and $\sin^22\theta$ for which the solar
neutrino data can be described at 95\% C.L. in terms of
$\nu_{e} \rightarrow \nu_{s}$ MSW transitions of the solar $\nu_e$
for values of ${\rm f_{Be}}$ equal to: a) 0.7; b) 1.0; c) 1.3, and for values
of ${\rm f_{B}}$ from the interval (0.40 -- 35). Shown in the figures is also 
the region excluded at 90\% C.L. by the absence of a difference between the 
day and the night event rates in the Kamiokande detector [12] (the region 
inside the dashed line). The plot for this excluded region was taken from
ref. [37].

\medskip

{\bf Fig. 2a.} Deformations of the spectrum of $^{8}$B ($\nu_e$) neutrinos
in the case of MSW transitions $\nu_e\rightarrow\nu_{s}$ of solar neutrinos 
for values of $\Delta m^2$ and $\sin^22\theta$ indicated in the figure. 
Each of the predicted $^{8}$B $\nu_e$ spectrum is divided by the standard
one and the ratio so obtained is normalized to the value it has at the
energy of $\nu_e$ E = 10 MeV (see, e.g., ref. [19]). The error-bars shown 
illustrate the envisaged sensitivity of the data from SNO (CC reaction)
after five years of operation; they do not include systematic errors.

\medskip

{\bf Fig. 2b.} Deformations of the spectrum of e$^{-}$ from the reaction
$\nu + e^{-} \rightarrow \nu + e^{-}$ caused by $^{8}$B neutrinos, in
the cases of MSW transitions $\nu_e\rightarrow\nu_{s}$ and for kinetic
energy of the electron ${\rm T_{e} \geq 5~ MeV}$.
%and $\nu_e\rightarrow\nu_{\mu(\tau)}$ (c). 
Each of the predicted recoil--electron spectrum is divided by the standard
one and the ratio so obtained is normalized to the value it has at 
T$_{e}=$ 10 MeV. The error-bars shown illustrate the sensitivity of the 
Kamiokande experiment [12] (large crosses) and the expected (ideal) sensitivity
of the Super--Kamiokande experiment after five years of operation 
(small crosses); in the latter case no systematic errors have been included.  

\medskip

{\bf Fig. 3a.} Deformations of the spectrum of pp ($\nu_e$) neutrinos
at ${\rm E \geq 0.15~MeV}$ in the case of the MSW $\nu_e\rightarrow\nu_{s}$ 
``low'' ${\rm f_{B}}$ solution (see the text) for $\sin^22\theta = 
5\times 10^{-3}$
and several values of $\Delta m^2$. Each of the predicted pp $\nu_e$ spectrum 
is divided by the standard one and the ratio so obtained is normalized to the 
value it has at the energy of the $\nu_e$ E = 0.3 MeV. The spectra shown in the
figure change very little when $\sin^22\theta$ is varied in the interval
$10^{-3} \ltap \sin^22\theta \ltap 2\times 10^{-2}$ (see Figs. 4a -- 4d).
%For the indicated values
%of $\Delta m^2$ the MSW $\nu_e\rightarrow\nu_{s}$ transitions lead to pp
%neutrino spectrum deformations very similar to the ones shown on the figure 
%also for $10^{-3} \ltap \sin^22\theta \ltap 10^{-2}$.

\medskip

{\bf Fig. 3b.} Deformations of the spectrum of pp ($\nu_e$) neutrinos
at ${\rm E \geq 0.15~MeV}$ in the case of the MSW $\nu_e\rightarrow\nu_{s}$ 
adiabatic solution (see the text). The spectra are normalized in the same way 
as those shown in Fig. 3a. 

\medskip 

{\bf Figs. 4a -- 4d.} Deformations of the spectrum of e$^{-}$ from the reaction
$\nu + e^{-} \rightarrow \nu + e^{-}$ caused by pp neutrinos, in
the case of MSW $\nu_e\rightarrow\nu_{s}$ transitions (nonadiabatic solution) 
for $\sin^22\theta = 10^{-3}~(a), 5\times 10^{-3}~(b), 10^{-2}~(c), 
2\times 10^{-2}~(d)$, at ${\rm T_{e} \geq 0.10~MeV}$. Each of the predicted 
recoil-electron spectra is 
divided by the standard one and the ratio so obtained is normalized to the 
value it has at the kinetic energy of the electron T$_{e}=$ 0.20 MeV. Note 
that the recoil-electron spectra change little with the change of 
$\sin^22\theta$.

\medskip

{\bf Fig. 5.} Deformations of the recoil-electron spectrum due to the MSW
$\nu_e\rightarrow\nu_{s}$ transitions (adiabatic solution) at 
${\rm T_{e} \geq 0.10~MeV}$ and for $\sin^22\theta = 0.30; 0.40$ and 
$\Delta m^2 = 6.0;~7.0;~8.0\times 10^{-6}~{\rm eV^2}$. The normalization
of the spectra is the same as in Figs. 4a -- 4d. In each of the three pairs of 
solid, dotted and dashed curves the upper (lower) curve at 
${\rm T_{e} < 0.20~MeV}$ (${\rm T_{e} > 0.20~MeV}$) corresponds to
$\sin^22\theta = 0.4$.

\end{document}